\documentclass[]{spie}  

\usepackage{amsmath,amsfonts,amssymb,mathrsfs}
\usepackage{graphicx,caption}
\usepackage[colorlinks=true, allcolors=blue]{hyperref}
\usepackage{gensymb}
\usepackage{mathtools}

\DeclarePairedDelimiter\abs{\lvert}{\rvert}%

\title{Multiplexed Holographic Aperture Masking with liquid-crystal geometric phase masks}
\author[a]{D. S. Doelman}
\author[b]{P. Tuthill}
\author[b,c]{B. Norris}
\author[a]{M. J. Wilby}
\author[a]{E.H. Por}
\author[a]{C.U. Keller}
\author[d]{M. J. Escuti}
\author[a]{F. Snik}

\affil[a]{Leiden Observatory, Leiden University, P.O. Box 9513, 2300 RA Leiden, The Netherlands}
\affil[b]{Australian Astronomical Observatory, 105 Delhi Rd, North Ryde NSW 2113, Australia}
\affil[c]{Sydney Institute for Astronomy (SIfA), Institute for Photonics and Optical Science (IPOS), Sydney Astrophotoinc Instrumentation Laboratory (SAIL), School of Physics, University of Sydney, NSW 2006, Australia}
\affil[d]{Department of Electrical and Computer Engineering, North Carolina State University, Raleigh, NC 27695, USA}

\authorinfo{E-mail: doelman@strw.leidenuniv.nl}

\begin{document} 
\maketitle

\begin{abstract}
Sparse Aperture Masking (SAM) allows for high-contrast imaging at small inner working angles, however the performance is limited by the small throughput and the number of baselines. We present the concept and first lab results of Holographic Aperture Masking (HAM) with extreme liquid-crystal geometric phase patterns. We multiplex subapertures using holographic techniques to combine the same subaperture in multiple non-redundant PSFs in combination with a non-interferometric reference spot. This way arbitrary subaperture combinations and PSF configurations can be realized, giving HAM more uv-coverage, better throughput and improved calibration as compared to SAM, at the cost of detector space.  
\end{abstract}


\section{INTRODUCTION}
\label{sec:intro}  
\begin{figure}[ht]
\center
\includegraphics[width = \textwidth]{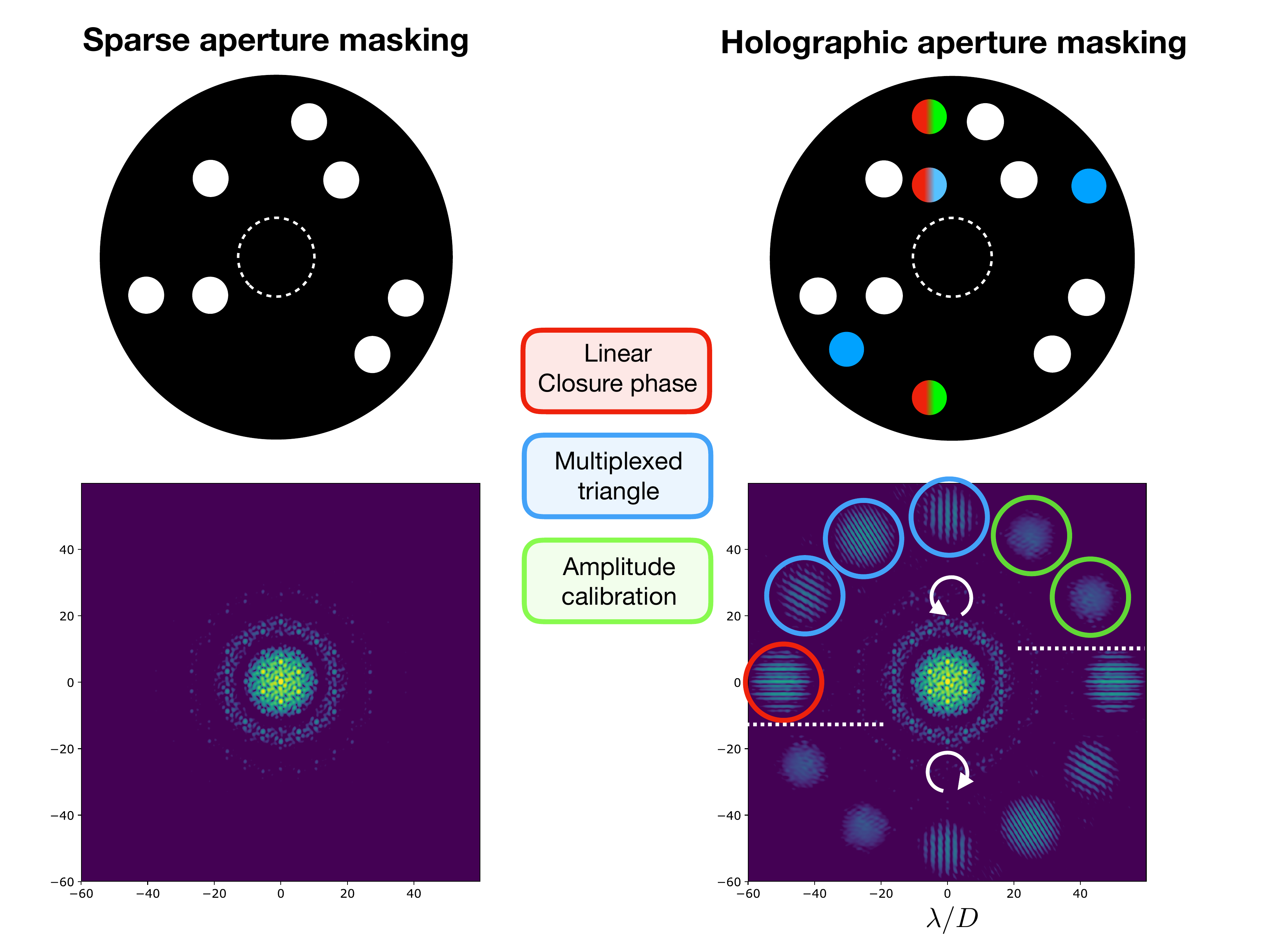}
\caption{Comparison of sparse aperture masking (left) and holographic aperture masking (right). Holographic aperture masking uses a phase plate to combine otherwise redundant subapertures at separate locations in the pupil plane. A non-redundant mask is shown on top left with the resulting PSF on the bottom left. Top right is a schematic of the HAM combinations, where each subaperture with the same color is combined at the PSF encircled in that color. Subapertures are multiplexed to make multiple PSF copies. Combinations can be one dimensional at the same location (red) or in triangles at a different location for each baseline (blue). In green are the two subapertures that have non-interferometric PSFs used for amplitude monitoring. Two copies of each PSF are created with opposite circular polarization state. 
 \label{fig:HAMSAM}}
\end{figure}

Sparse aperture masking (SAM) is a technique that turns a single dish telescope into an interferometer by masking out a large fraction of the pupil. The opaque mask consists of a sparse combination of holes such that the point-spread function (PSF) is a combination of interferometric fringes from each baseline. Because SAM is an interferometric method it is possible to measure down to half the diffraction limit of a single dish telescope. More importantly, with SAM it is possible to measure closure phases, an observable that is independent of the incoming wavefront aberration. Both advantages make SAM a good option for high contrast imaging and with the improved calibration that came with adaptive optics systems, sparse aperture masking has been extremely successful in imaging asymmetric structures around stars at small separations unreachable by other techniques like coronagraphy. SAM has been used to measure the shape and grain sizes of dust shells around stars \cite{Norris2012}, discovering substellar companions around young stars \cite{Kraus2012,Huelamo2011} and measure stellar multiplicity in star-forming regions \cite{Ireland2008,Martinache2009,Cheetham2015}. \\
Non-redundant masking is a subtechnique of SAM and requires the holes to be places in non-redundant patterns. Only a limited amount of holes can be combined in a non-redundant way and non-redundant masks therefore have a low throughput. Techniques like segment tilting and pupil remapping are ways to improve the throughput by making different non-redundant combinations of the aperture. Both techniques are complex to implement for a given telescope compared to the simplicity of sparse aperture masking where only one mask with holes is needed. Holographic aperture masking (HAM) uses a single phase plate to combine non-redundant subapertures at different focal plane locations, maintaining the simplicity of SAM. The possibilities with HAM go beyond the segment tilting as holographic techniques can be used to make multiple copies of each subaperture, enabled by liquid-crystal technology. A comparison between SAM and HAM is given in Fig. \ref{fig:HAMSAM}. HAM can incorporate SAM while adding more baselines by interfering additional subapertures with PSF copies at a separate location in the focal plane. While more detector space is required, HAM has more baselines and closure phases, uses more subapertures, allows for broadband operation and can be used to generate amplitude reference spots. \\
In this paper, we present the theory for blazed gratings and liquid-crystal technology in section \ref{sect:theory}, the design of holographic aperture masks in section \ref{sect:design}, lab results in section \ref{sect:lab} and conclusions are presented in section \ref{sect:conclusion}.

\section{Theory}
\label{sect:theory}
Holographic aperture masking (HAM) uses the freedom that any subaperture of the pupil can be imaged on any place in the focal plane. These subaperture spots are interfered with other subapertures by placing copies of the point-spread function (PSF) of both apertures on the same location. This allows for arbitrary combinations of subapertures at any location of the focal plane. We generate these copies of a subaperture PSF using holographic blazed gratings. 
\subsection{Multiplexed holographic blazed gratings}
\begin{figure}[ht]
\center
\includegraphics[width = \textwidth]{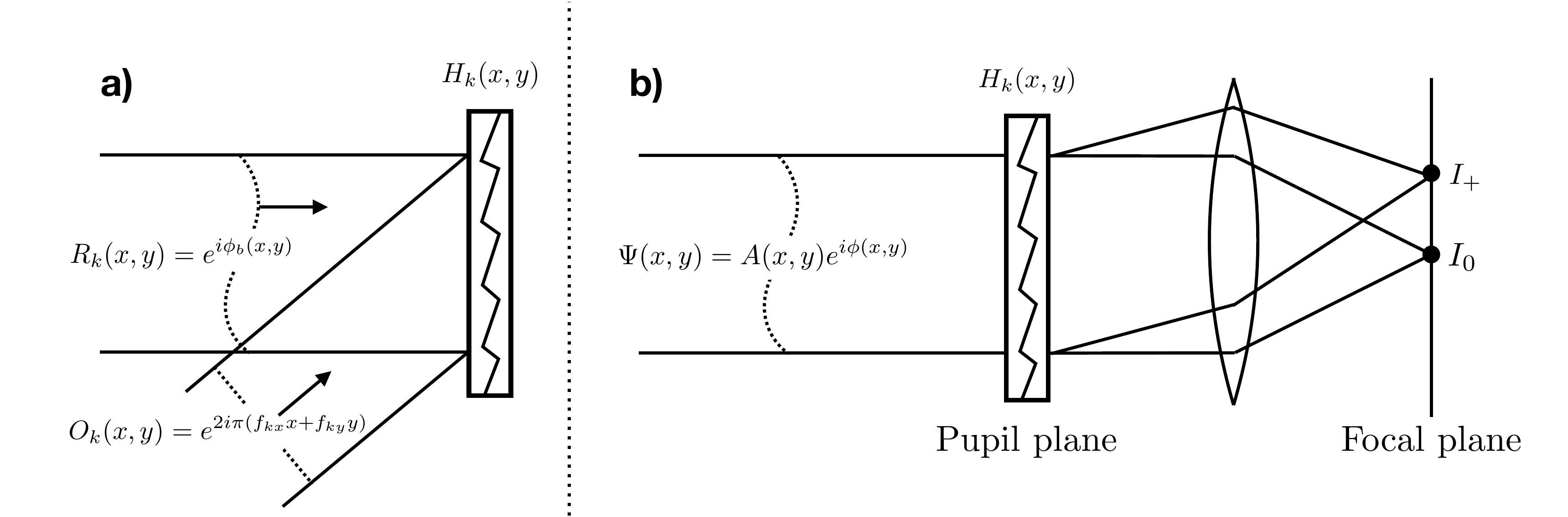}
\caption{Diagram of holographic blazed gratings. \textbf{a)} Visualized generation of the holographic phase pattern. $H_k(x,y)$ is blazed by using only one interference term. \textbf{b)} A selected amount of the incoming wavefront is imaged at a separate location in the focal plane. Image adapted from Dong et al. (2012) \cite{Dong2012} and Wilby et al. (2016)\cite{Wilby2016}.
 \label{fig:genholograms}}
\end{figure}

Blazed gratings have an optimal diffraction efficiency to a single order. The generation of these holograms is depicted in Fig. \ref{fig:genholograms}, where we adapt the notation of Dong et al. \cite{Dong2012}. 

We generate an interferogram between a reference wavefront $R_k(x,y)$ with a biased phase $\phi_b(x,y)$ and a object wavefront $O_k(x,y)$. The reference wavefront is given by
\begin{equation}
R_k(x,y) = e^{i\phi_b(x,y)}
\end{equation}
and the object wavefront by
\begin{equation}
O_k(x,y) = e^{2 i \pi (f_{kx}x + f_{ky}y)},
\end{equation}
where $f_{kx}$ and $f_{ky}$ are the spatial frequencies the hologram is placed in the focal plane, $f_{kx} = x^\prime_k/f\lambda$. Here the focal plane coordinates are given by $(x^\prime_k,y^\prime_k)$. HAM does not require a biased reference wavefront other than a piston term ($\phi_b(x,y) = c_k$) that is used to phase scramble interferometric PSFs. The interferogram $H_k(x,y)$ between the two waves is then given by
\begin{equation}
H_k(x,y) = \abs{O_k(x,y) + R_k(x,y)}^2
\end{equation}
\begin{equation}
H_k(x,y) = \abs{O_k}^2 + \abs{R_k}^2 + O^*_k R_k + O_k R^*_k
\end{equation}
\begin{equation}
H_k(x,y) = 2 + O^*_k R_k + O_k R^*_k,
\label{eq:interferogram}
\end{equation}
where $^*$ stands for the complex conjugate operator. The interferogram now generates two PSF copies, the $\pm 1$ orders of the grating. Having only one of the two copies is preferred for HAM, as having two would increase the necessary detector space by a factor of two. We therefore blaze the grating by selecting one interference term, 
\begin{equation}
H_k(x,y) = O_k R^*_k, 
\end{equation}
and only one PSF copy is created. Selecting one term is possible because we numerically generate the hologram. \\
The complex electric field in the pupil entering the holographic blazed grating can be described by 
\begin{equation}
\Psi(x,y) = A(x,y) e^{i\phi(x,y)},
\end{equation}
where $A(x,y)$ is the pupil function of a subaperture of the telescope and $\phi(x,y)$ is the incoming wavefront. The consecutive focal plane intensity is given by $I = \abs{\mathscr{F}\left[H_k\Psi\right]}^2$. Assuming a blazed grating, the intensity is given by
\begin{equation}
I_k(x^\prime_k,y^\prime_k) = \delta(x^\prime - x^\prime_k)\delta(y^\prime - y^\prime_k) * \abs{\mathscr{F}\left[A(x,y)\right]}^2 * \left|\mathscr{F}\left[e^{i(\phi(x,y)+c_k)}\right]\right|^2,
\label{eq:PSF}
\end{equation}
using $*$ is the convolution operator. Eq. \ref{eq:PSF} is the subaperture PSF at $(x^\prime_k,y^\prime_k)$ with a phase offset $c_k$. The blazed grating therefore only generates one PSF copy.  
\\
Changing the PSF locations of subapertures is also possible with mirrors \cite{tuthill2012unlikely}. Holography, however, allows for much more freedom than beam tilting only. An example of the freedom is phase scrambling and was already mentioned above. More importantly, holography enables multiplexing of holograms, creating more than one copy for each subaperture. HAM does benefit greatly from this property, we can choose to interfere subapertures arbitrarily with selected intensities anywhere in the focal plane. We multiplex blazed gratings by taking the sum over all complex interferograms scaled by a factor $s_k$. Both the reference and object wavefront have unity amplitude and therefore we create a phase-only hologram by taking the argument of the multiplexed hologram,
\begin{equation}
\phi_h(x,y) = \frac{1}{\pi} \text{arg} \left[ \sum_k^N s_k H_k(x,y) \right].
\end{equation}

\begin{figure} [ht]
\center
\includegraphics[height=10cm]{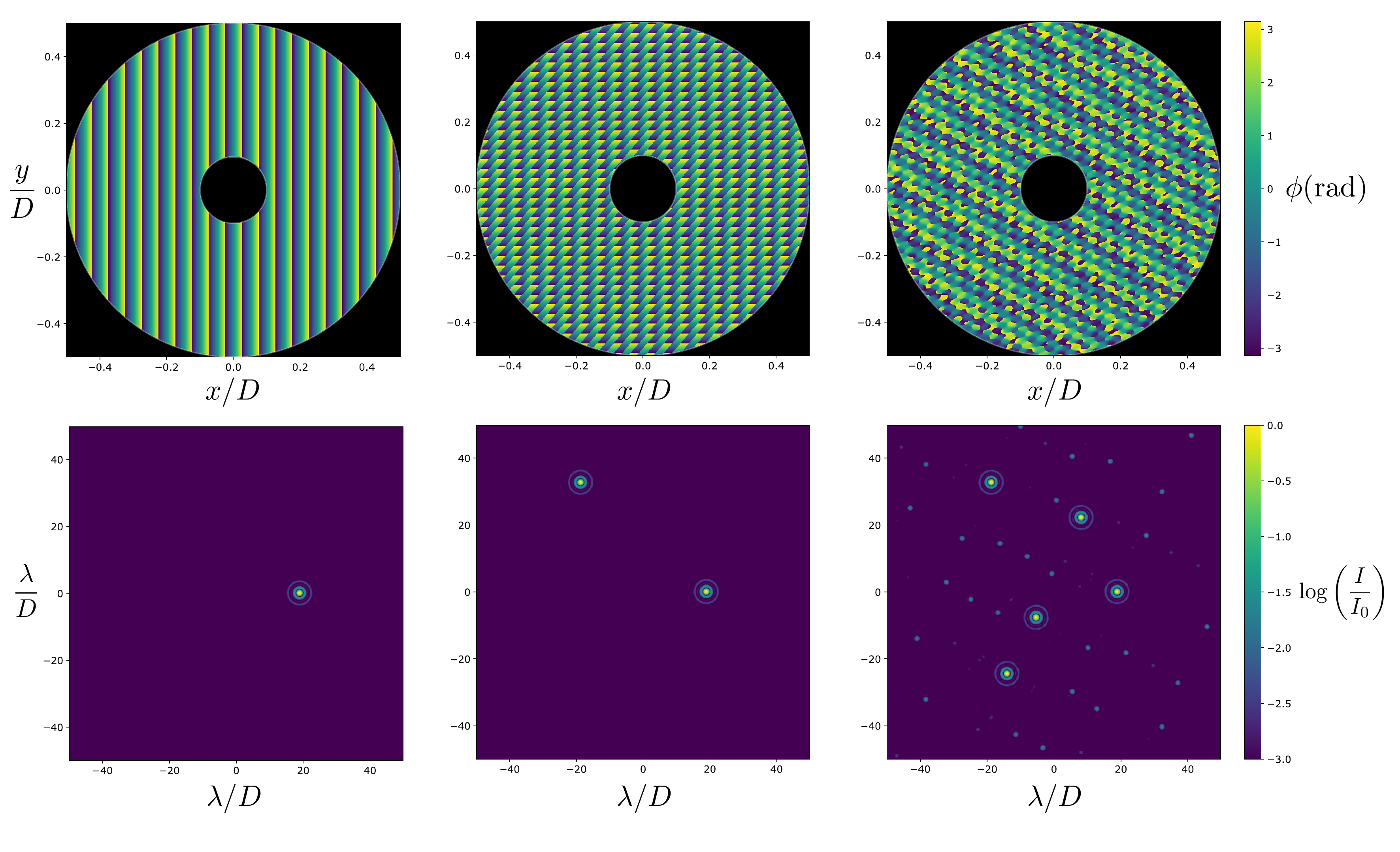}
\caption{Holographic multiplexing of an aperture. The light is multiplexed in one, two and five holograms from left to right respectively. 
\label{fig:multiholo}}
\end{figure}

An example of a multiplexed holographic grating is shown in Fig. \ref{fig:multiholo}. 

\subsection{Liquid-crystal technology}
Phase patterns with the complexity required for HAM can be generated by applying geometric phase (or Pancharatnam-Berry) phase to circularly polarized light. When light that is left- or right-handed circularly polarized propagates through a half-wave retarder, the light acquires not only phase from the optical path difference between the fast-and slow-axis of the retarder, but also an extra phase shift is introduced. This extra phase shift $\phi$ only depends on the orientation of the fast-axis $\theta$ of the half-wave retarder and the circular polarization state \cite{pancharatnam1956generalized,Berry1987,Escuti:16} 
\begin{equation}
\phi = \pm 2 \theta.
\label{eq:signphase}
\end{equation}
Note that the geometric phase is independent of wavelength but requires a retardance that is half-wave. When the retardance deviates from half-wave, only the amount of light that acquires the geometric phase (=efficiency), decreases and a leakage term emerges. The leakage term does not acquire any geometric phase. \\
Generating the complex phase pattern for HAM requires an optic with locally varying fast-axis orientation and a half-wave retardance. For broadband operation the retardance needs to be tuned to the desired wavelength range. Both requirements can be satisfied with liquid-crystal technology. A fast-axis orientation pattern can be written in a liquid-crystal polymerizable polymer (LCP) layer using a direct-write method \cite{Miskiewicz2014}. This photo-alignment layer orients itself to the varying angle of linear polarization of the incoming uv-laser that scans the substrate. The retardance is tuned using birefringent self-aligning liquid-crystal layers. By changing the thickness and twist in each layer it is possible to achieve a retardance close to half-wave for very large bandwidths ($>100\%$)\cite{Komanduri2013}. The layers are cured with uv-radiation, such that the device passively retains it broadband performance. Many complex patterns have been manufactured with this technology already \cite{Escuti:16,Kim2015,Doelman2017PatternedSensing}. 

\subsection{Polarimetry with HAM (HAMPol)}
Sparse aperture masking can be combined with polarimetry (SAMPol), making optimal use of the very small inner working angles with the ability to measure polarization signals \cite{tuthill2010sparse}. These polarization signals usually arise from scattering or reflection of light and SAMPol was used to spatially resolve circumstellar dust shells at less than 2 stellar radii of three stars on the asymptotic giant branch \cite{Norris2012}. SAMPol is complementary to other techniques because SAMPol probes the most inner stellar regions not resolvable by conventional imaging polarimetry. The VAMPIRES instrument installed at the Subaru telescope takes advantage of the small diffraction limit in the visible to eventually scales as small as 10mas \cite{Norris2015}.\\
With holographic aperture masking it is also possible to measure polarization, similar to how the vector-apodizing phase plate can be used for polarimetry \cite{snik2012}. The thin liquid-crystal film applies geometric phase to circular polarization states. The applied phase has a different sign for both polarization states (Eq. \ref{eq:signphase}). Combined with the ability to make single holographic copies, the detector will have two holographic PSFs at opposite locations, one for each circular polarization state. When the liquid-crystal device is placed between quarter-wave plates, HAM separates linear polarization. HAM is therefore a natural implementation polarimetric imaging with aperture masking.  

\section{Design}
\label{sect:design}
There is a lot more freedom when designing a holographic aperture mask compared with a sparse aperture mask. Designing a sparse aperture mask involves optimizing the number of apertures, aperture sizes and their location to maximize uv-coverage and throughput under the constraint that they are non-redundant \cite{carlotti2010new}. The advantage of HAM is that it allows to multiplex any combination of subapertures at any location in the focal plane. The throughput of HAM is then easily increased compared to SAM by imaging multiple non-redundant combinations at different locations. The non-redundancy constraint still applies to combinations at the same focal plane location but not to combinations at different locations. In addition, subapertures can be multiplexed and can therefore be used more than once with any relative intensity. The signal to noise in the focal plane is optimal when combinations of subapertures have the same intensity, but that is not constrained by the technology. In the end, HAM is limited to the number of non-redundant combination PSFs that fit the detector space, as every PSF has the size of $\lambda/D_{s}$, where $D_{s}$ is the diameter of the subaperture. 

\subsection{Bandwidth}
HAM manufactured with liquid-crystal technology applies geometric phase that is independent of wavelength. While this would allow for broadband operation, the element is diffractive and wavelength smearing washes out fringes quickly for off-axis PSFs. Mirror tilting does not have this disadvantage because it applies optical path difference (OPD), such that the center of the off-axis PSFs does not change with wavelength \cite{tuthill2012unlikely}. A holographic grating that places a PSF copy at $30 \lambda/D$ smears the PSF over $0.3 \lambda/D$ for a $1\%$ bandwidth. Any gain in throughput from HAM is then nulled by limiting the bandwidth to $1\%$. There are two solutions to this problem. The first one is to use the design freedom to combine subapertures in only one direction such that the fringes are one dimensional and can be placed orthogonal to the smearing direction. The second solution is to use a Wynne lens that corrects the dispersion by introducing a chromatic difference of magnification with opposite sign. This is not entirely similar to segment tilting as off-axis sources are still dispersed and the dispersion is a function of the distance to the star. Using a Wynne lens has the advantage that it does increase the signal per pixel, however it consists of a set of custom lenses and the dispersed planet changes the fringe pattern. 

\subsection{One dimensional combinations}
\begin{figure} [ht]
\center
\includegraphics[height=8cm]{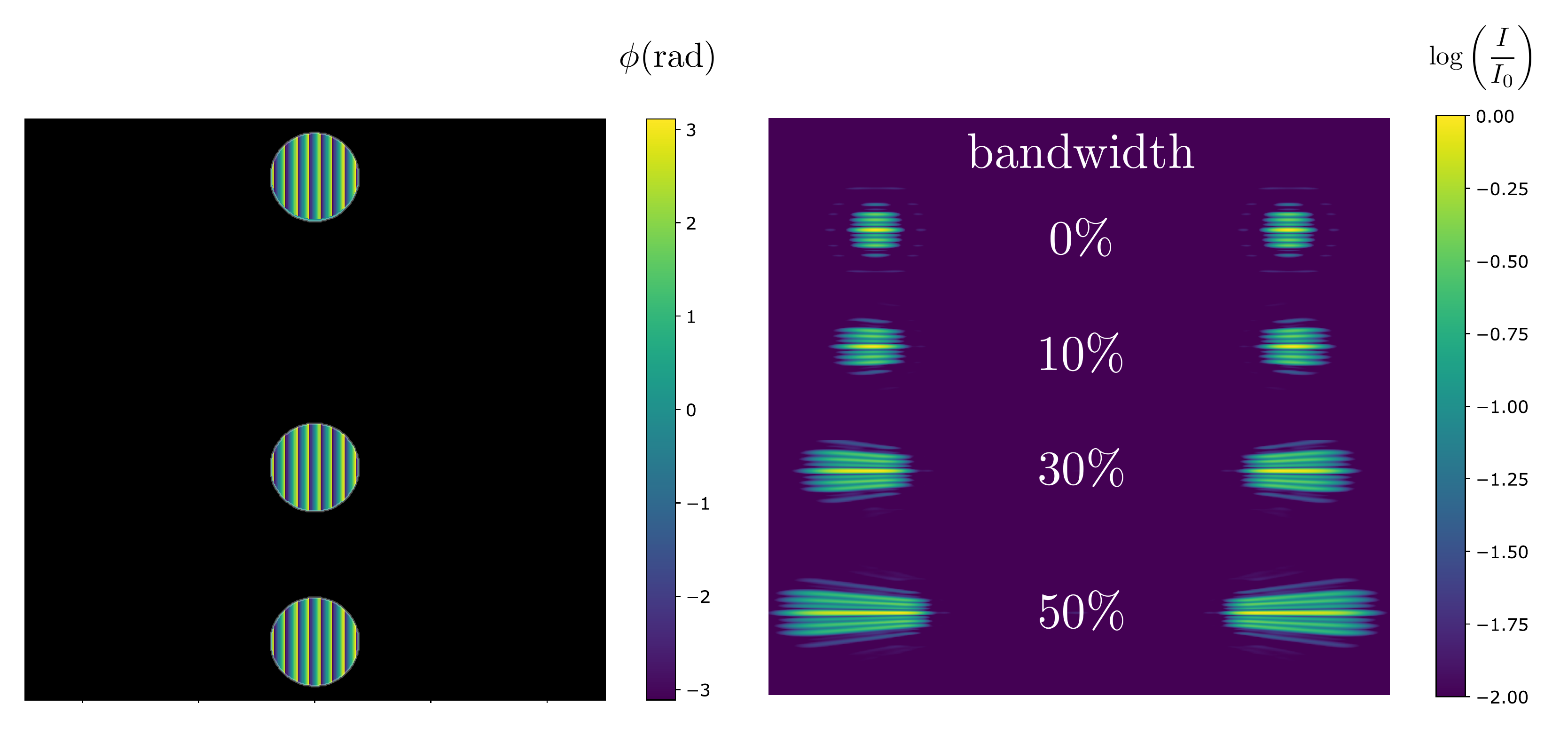}
\caption{Bandwidth smearing of fringes (right) from a one dimensional combination of subapertures (left). 
 \label{fig:bwsmearing}}
\end{figure}

Combinations of subapertures in one dimension will create fringes in the same direction. We place these fringes orthogonal to the wavelength smearing direction such that we can increase the bandwidth arbitrarily without decorrelating the fringes. This is seen in Fig. \ref{fig:bwsmearing}. The fringes appear on both sides for opposite polarization states, no leakage was simulated. 
Using achromatic phase enabled by liquid-crystal technology, together with the optimal fringe placement, HAM is able to operate on bandwidths much larger than a few percent that is typical for sparse aperture masking. HAM design is now a trade-off between bandwidth and detector real estate. \\

Only combining apertures in one dimension at one point in the focal plane greatly reduces the number of baselines and closure triangles. The larger bandwidth and the ability to use more subapertures can in certain cases make up for the loss in the number of closure phases, especially when the detector is large and the observed object has a simple geometry. However, closure phases are the most used way of acquiring unambiguous measurements of an observed object. Another way to increase the number of baselines and closure phases is to add holograms to a SAM design, or some light can be multiplexed from a SAM design to locations on the detector for better amplitude calibrations, as shown in Fig. \ref{fig:HAMSAM}. The bandwidth is then limited by the SAM design.\\
Having a large bandwidth is not only useful for more photons on the detector. The phase offsets of each subaperture is wavelength dependent and therefore the fringes will change as function of wavelength. Fitting the broadband fringes could enable the extraction of closure phases, increasing the amount of information stored in a single hologram. This was not simulated and is outside the scope of this paper.

\subsection{Multiplexed gratings}
When the number of combined subapertures in one dimension is larger than two, it is possible to reconstruct a closure phase and there is no need for multiplexing the subapertures. Multiplexing is useful for combining subapertures that do not line up. A closure triangle can be obtained by combining the individual combinations per two subapertures at the location orthogonal to that specific baseline. This way, the bandwidth smearing is radial and a much larger bandwidth can be used compared to combining the three subapertures on one location of the focal plane. Multiplexing can also be used to link two separate subaperture combinations. However, to keep the intensities of all individual subapertures combinations the same, the percentage of light per holographic PSF needs to be scaled. As an example we assume that subapertures A, B and C are interfered at three different locations in the focal plane and A is linked for $40\%$ to another combination D. Now A only has $60\%$ of the light to make the combinations with B and C. Then the A-B combination with equal intensity uses only $30\%$ from subaperture B. This can be compensated by making the B-C combination use $70\%$ of subaperture B and C. From this example it is clear that optimizing the combinations with multiplexing can be very complex. \\
Moreover, multiplexing individual subapertures can be used for amplitude calibration by making a copy that is not interfered with any of the other subapertures. By measuring the encircled energy of this hologram, assuming the subaperture size is on the size of the Fried parameter $r_0$, the amplitude is directly obtained and no longer a fit parameter in fringe fitting. This could improve the precision of the calculated closure phases.
\begin{figure}[ht]
  \centering
   \captionsetup{width=.9\linewidth}
  \includegraphics[width=\linewidth]{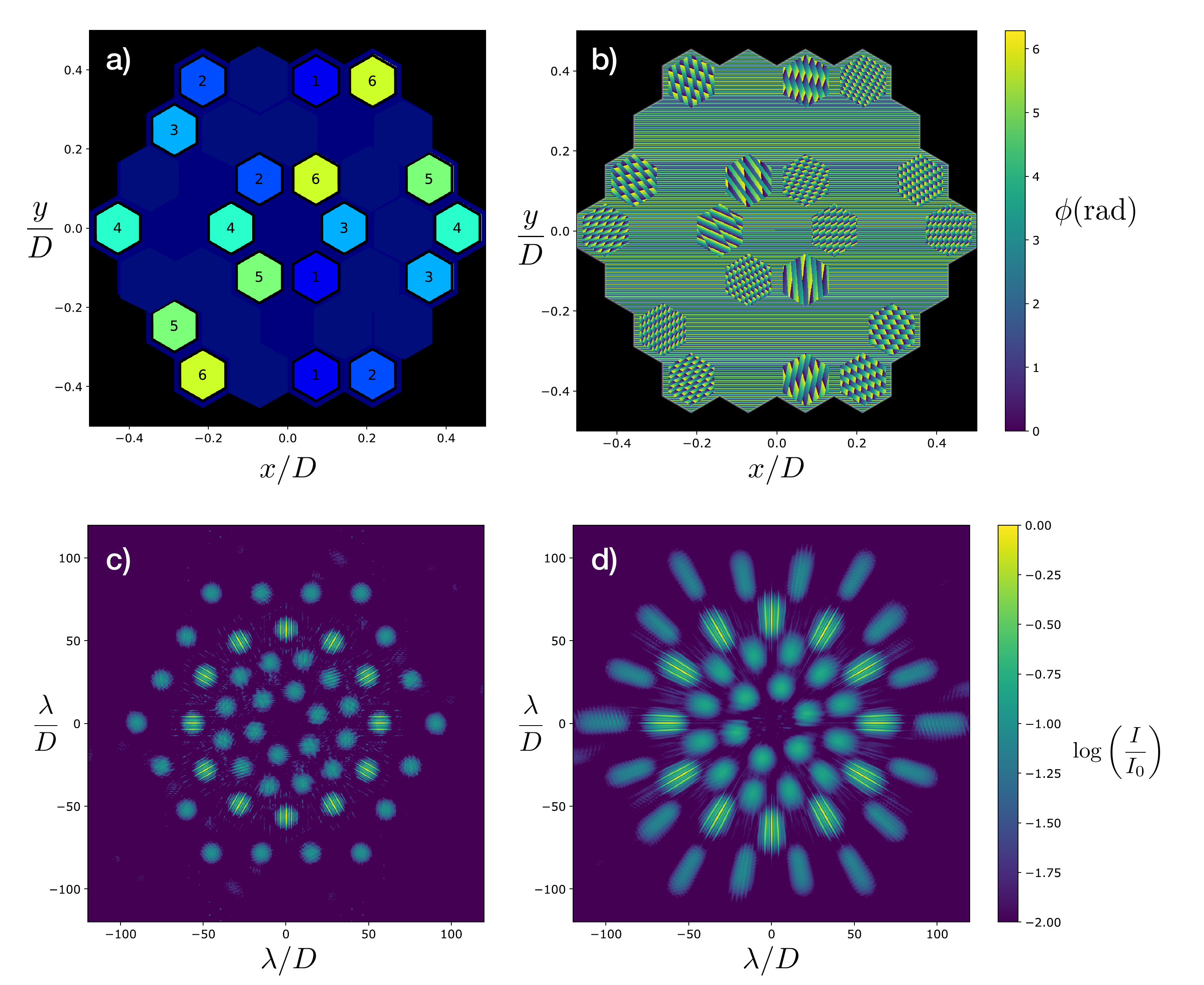}
  \caption{\textbf{a)} Diagram with the combinations of subapertures. Every subaperture with the same number is combined in one PSF on the focal plane and every subaperture is multiplexed to have a non-interferometric reference PSF. \textbf{b)} Phase pattern that corresponds to the design. The subapertures are cut out using a grating mask. \textbf{c)} Simulated PSF of the design, where all interferometric PSFs are placed in a ring. The reference PSFs within this ring are placed in smaller concentric rings, while the reference PSFs outside the ring are placed on a hexagonal grid. \textbf{d)} Simulated PSF for $30\%$ broadband light. No indiviual PSFs overlap.    
  \label{fig:hamdesign1}}
\end{figure}
\section{Lab results}
\label{sect:lab}
We designed two holographic aperture masks to be tested in the lab. We used a hexagonal aperture from Keck and the subapertures are similar to undersized single mirror segments. Using such a regular grid and relatively large subapertures greatly reduces the degrees of freedom. The designs are optimized by hand as a global optimization process is still being researched. By printing a grating mask, a high frequency grating outside of the used subapertures, there is no need for a real amplitude mask for the specific designs \cite{Doelman2017PatternedSensing}. 
\begin{figure}
  \centering
   \captionsetup{width=.9\linewidth}
  \includegraphics[width=\linewidth]{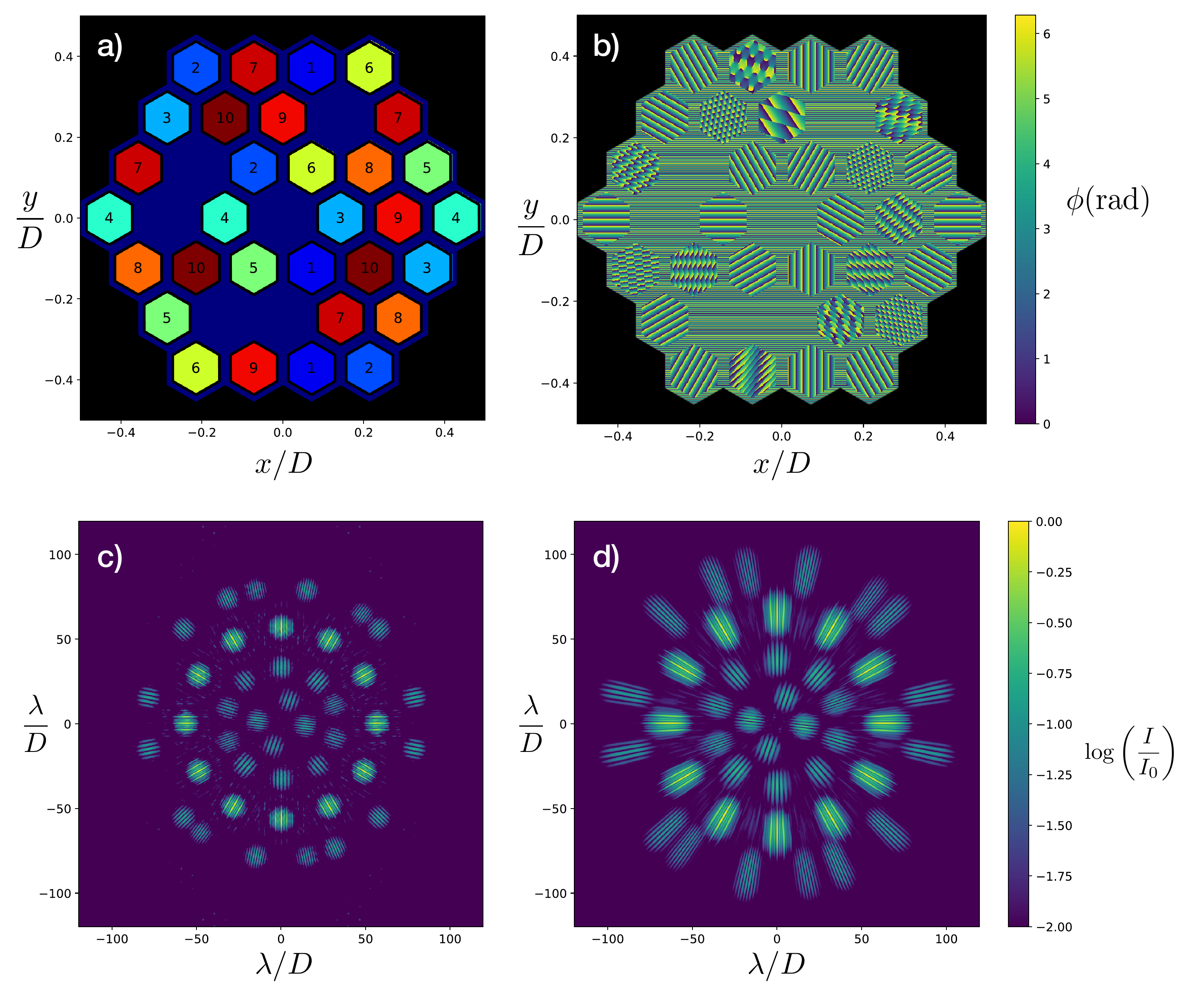}
  \caption{\textbf{a)} Diagram with the combinations of subapertures. Every subaperture with the same number is combined in one PSF on the focal plane. The pattern is optimized for the number of subapertures and closure phases. \textbf{b)} Phase pattern that corresponds to the design. The subapertures are cut out using a grating mask. \textbf{c)} Simulated PSF of the design, where all interferometric PSFs are placed in concentric rings. \textbf{d)} Simulated PSF for $30\%$ broadband light. No individual PSFs overlap.   
  \label{fig:hamdesign2}}
\end{figure}
\subsection{Manufactured design}
The first design combines all possible one dimensional combinations of three subapertures. The numbering and color in panel \textbf{a)} indicate the which subapertures are combined at one location of the focal plane. For this design $18/36$ subapertures are used to get 18 baselines and 6 closure phases. We multiplexed every subaperture such that it has a copy that can be used for amplitude calibration. The resulting phase pattern is shown in panel \textbf{b)}. The monochromatic PSF and the PSF for $30\%$ bandwidth are shown in panel \textbf{c)} and \textbf{d)} respectively. The total amount of spots is 48. The six interference point-spread functions are combined at $60 
\lambda/D$ to minimize crosstalk and to keep the necessary field of view as small as possible. The less important individual PSFs have been added to form the most compact configuration, allowing for some crosstalk.\\
The second design has the same optimal basis as the first design with all combinations of three subapertures. We optimized the amount of used subapertures, baselines and closure phases by hand, starting from the interferometric PSFs of design one. The second design has an extra combination of four subapertures, number 7 in Fig. \ref{fig:hamdesign2}  \textbf{a)}, and extra combinations of three subapertures, number 8-10. In total 31/36 possible subapertures have been used for 33 baselines and 12 closure phases with 42 spots. The combinations, the phase and the PSFs are displayed in  Fig. \ref{fig:hamdesign2}. 

\subsection{Manufacturing and lab testing}
\begin{figure} [ht]
\center
\includegraphics[width=0.6\linewidth]{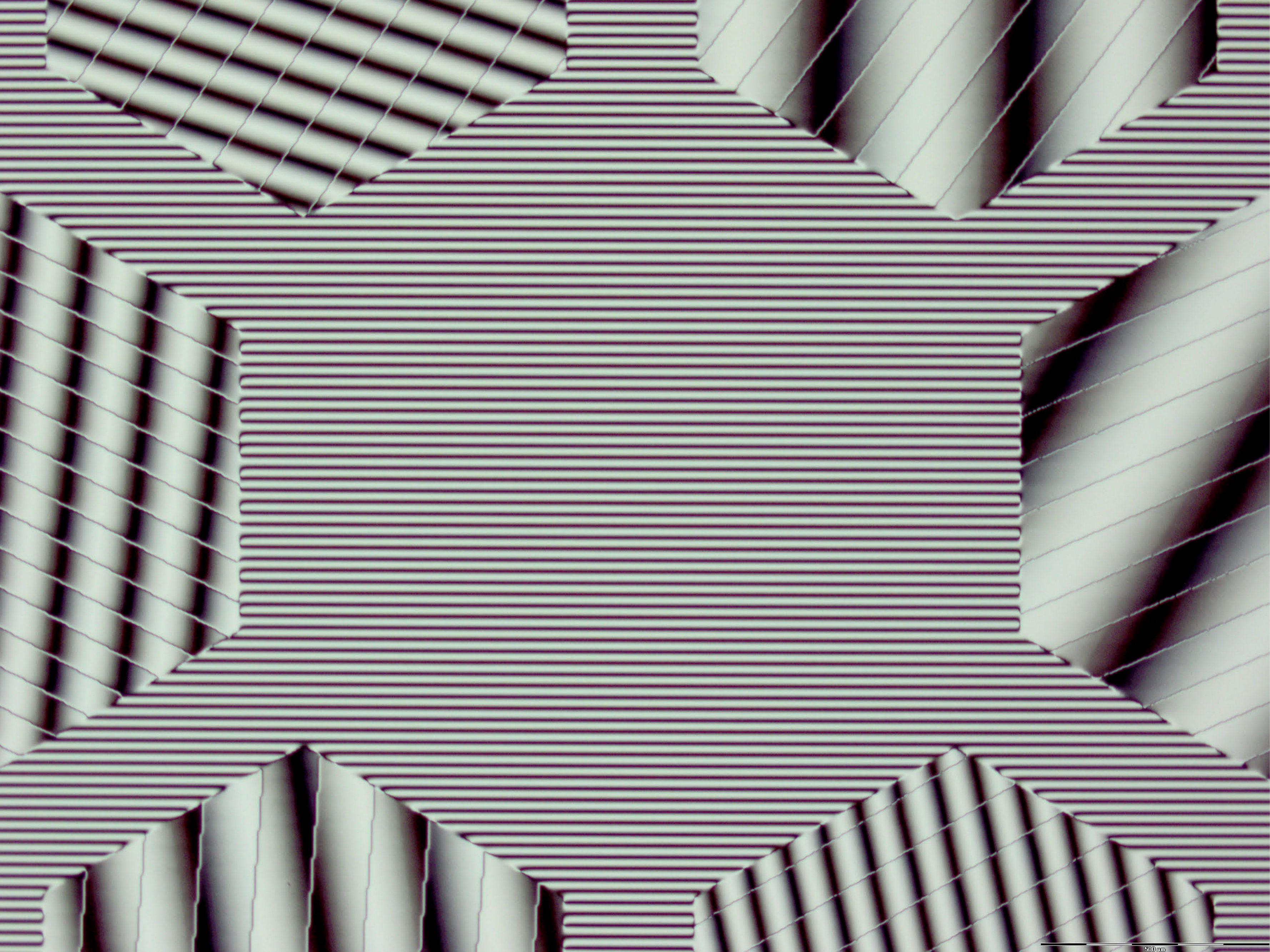}
\caption{Microscopic image of the liquid-crystal pattern for HAM design one through crossed polarizers. The image is magnified 20 times and only contains the central part. The pattern matches the design to a very high degree.
 \label{fig:crossed_pol}}
\end{figure}

\begin{figure}[ht]
\center
\includegraphics[trim={0 1.3cm 0 1.3cm},clip,width=0.75\linewidth]{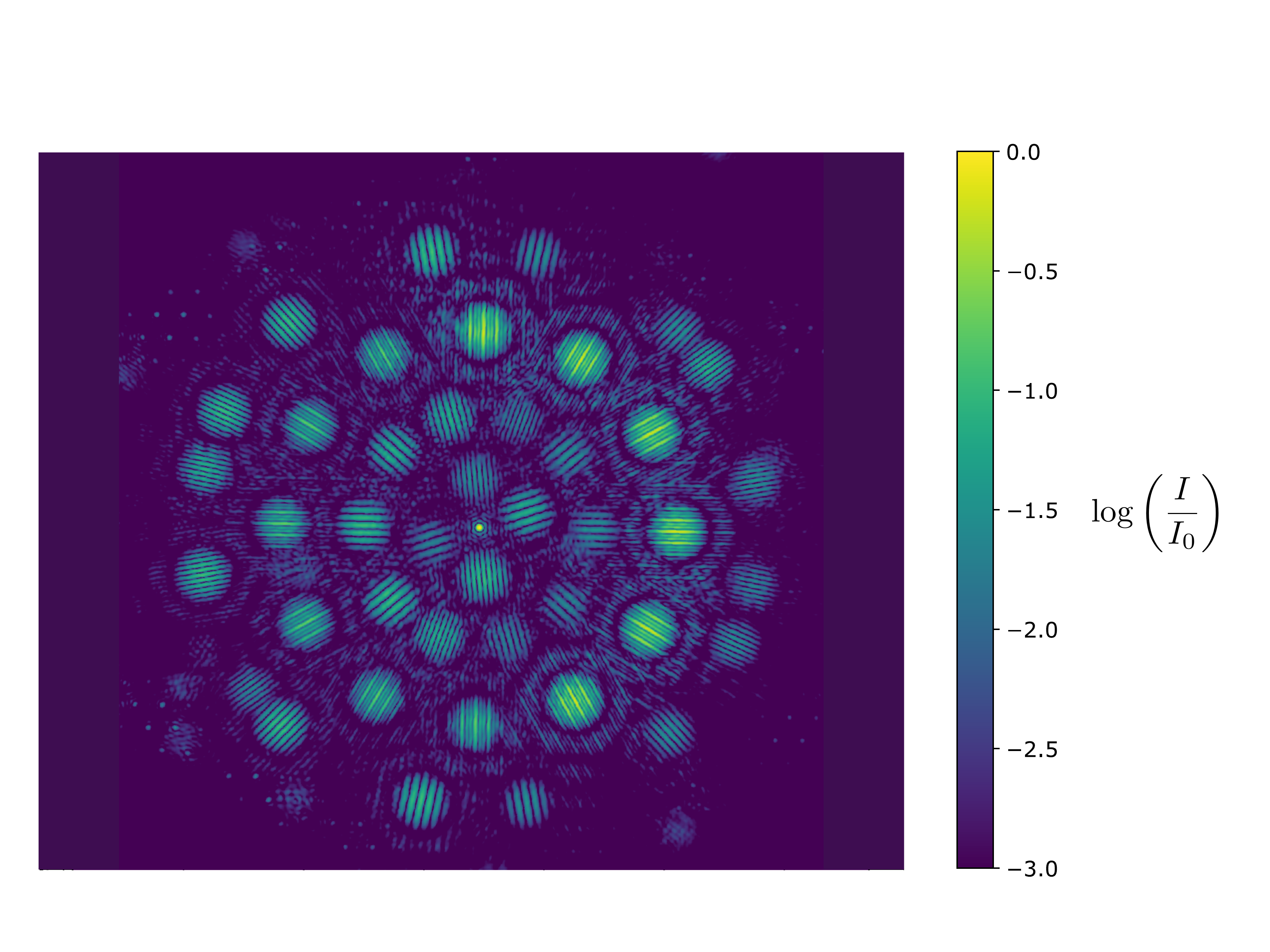}
\caption{Simulated PSF for HAM design 2 for monochromatic light and a 1:3 ratio between left and right circular polarization. The simulated amount of leakage is $0.2\%$. 
 \label{fig:SIM_PSF}}
\end{figure}

\begin{figure}[ht]
\center
\includegraphics[trim={0 1.3cm 0 1.3cm},clip,width=0.75\linewidth]{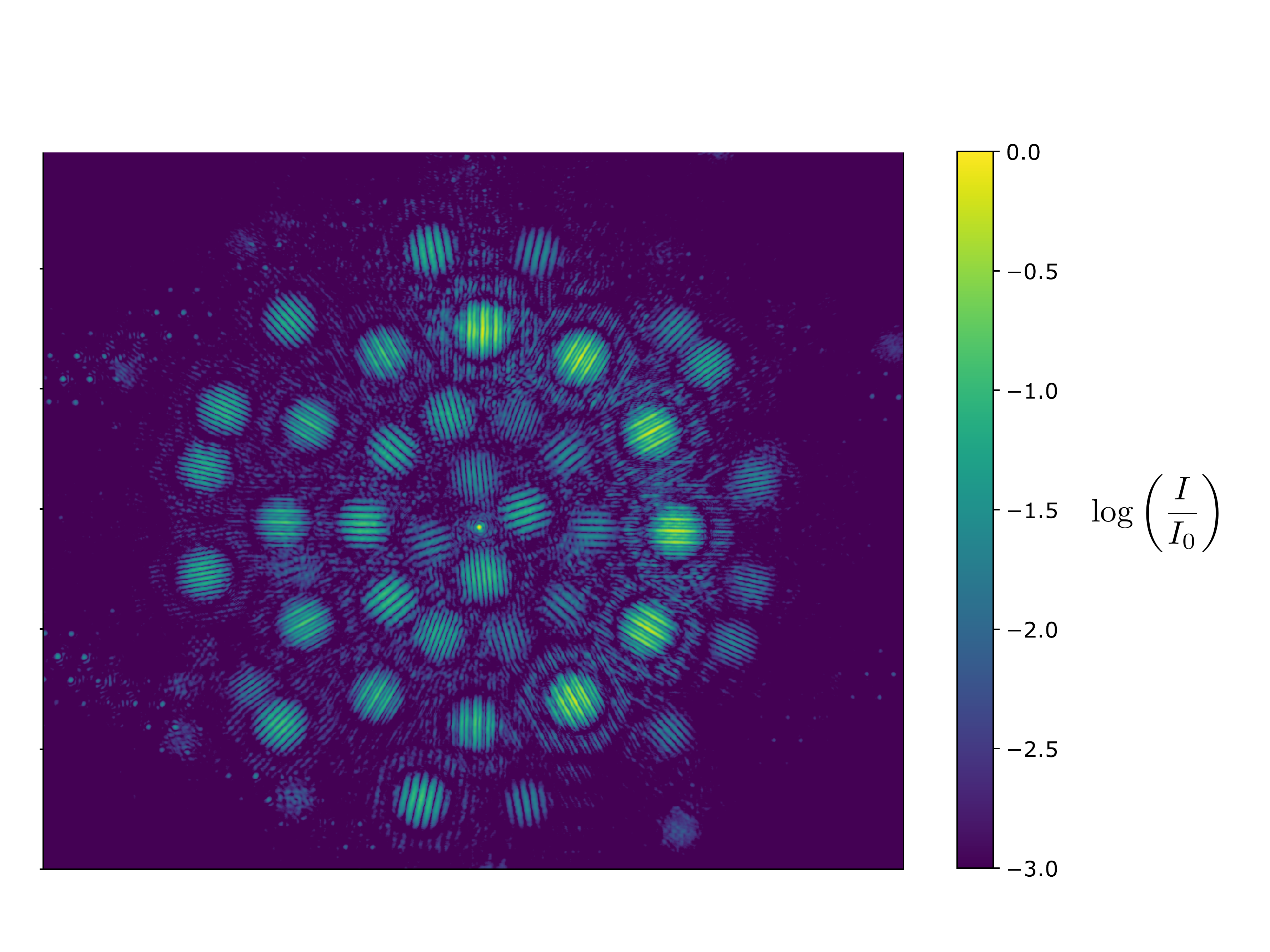}
\caption{Measured PSF for HAM design 2 at a wavelength of 532nm.
 \label{fig:LAB_PSF}}
\end{figure}

The designs were manufactured in December 2017 and both were printed on the same two inch substrate with a single layer of liquid-crystals. The single layer has an optimized thickness to tune the retardance to be half-wave at a wavelength of $532$nm. The pixel size of the pattern is 5 micron and the maximum width of the hexagonal aperture is 11mm. The subapertures are undersized by a factor of \textbf{x} and the grating period of the grating mask is \textbf{35 micron}. No anti-reflection coating or wedges have been added. The pattern was inspected between crossed polarizers under a microscope with a magnification of 20. The resulting image is shown in Fig. \ref{fig:crossed_pol}. No significant differences between the design and the manufactured pattern were found. 
We tested the lab plates on a preliminary version of the Leiden HCI testbed. A laser diode (CPS532) from Thorlabs operating at 532nm is fed into a single mode fiber for a diffraction limited input. The output of the fiber is collimated and an iris stops the pupil down to a size of 12mm. The holographic aperture mask is conjugated to this pupil plane with two intermediate focii and one reflective pupil where no active element is installed. We image the beam on a SBIG detector with a field of view of more than 200x200 $\lambda/D$. We compare the result with a simulated PSF where the ratio between left and right circular polarization was matched to the measurement. The simulated PSF is shown in Fig. \ref{fig:SIM_PSF} and the measured PSF is shown in Fig. \ref{fig:LAB_PSF}. The measured PSF has only very small deviations from the simulated PSF, showing that liquid-crystal technology is capable of accurately making the phase patterns required for HAM. 
\newpage

\newpage

\section{Conclusion}
\label{sect:conclusion}
\itemize{
\item Liquid-crystal technology enables the use of holographic techniques for masking interferometry.
\item A standalone holographic aperture mask (HAM) with one dimensional subaperture combinations can operate up to more than $30\%$ bandwidth, however the number of closure phases is limited.
\item We spectrally resolve fringes with HAM using these one dimensional subaperture combinations with large bandwidths. 
\item Multiplexing subapertures not used by sparse aperture masking (SAM) increases the number of baselines, closure phases and throughput at the cost of detector space.
\item Leakage from deviations of half-wave retardance interfere with the SAM point-spread function but can be controlled using a double grating technique. 
\item Holographic aperture masking works in the lab and will be on sky soon.
\item Liquid-crystal technology also can be used for broadband achromatic nullers. 
}

\acknowledgments 
The research of David S. Doelman and Frans Snik leading to these results has received funding from the European Research Council under ERC Starting Grant agreement 678194 (FALCONER)
{\scriptsize
\bibliography{Mendeley}} 

\begin{thebibliography}{10}

\bibitem{Norris2012}
Norris, B.~R., Tuthill, P.~G., Ireland, M.~J., Lacour, S., Zijlstra, A.~A.,
  Lykou, F., Evans, T.~M., Stewart, P., and Bedding, T.~R., ``{A close halo of
  large transparent grains around extreme red giant stars},'' {\em Nature}~{\bf
  484}(7393),  220--222 (2012).

\bibitem{Kraus2012}
Kraus, A.~L. and Ireland, M.~J., ``{LkCa15: A young exoplanet caught at
  formation?},'' {\em Astrophysical Journal}~{\bf 745}(1) (2012).

\bibitem{Huelamo2011}
Huelamo, N., Lacour, S., Tuthill, P., Ireland, M., Kraus, A., and Chauvin, G.,
  ``{A companion candidate in the gap of the T Cha transitional disk},'' ~{\bf
  7},  2009--2012 (2011).

\bibitem{Ireland2008}
Ireland, M.~J., Kraus, A., Martinache, F., Lloyd, J.~P., and Tuthill, P.~G.,
  ``{Dynamical Mass of GJ 802B: A Brown Dwarf in a Triple System},'' {\em The
  Astrophysical Journal}~{\bf 678}(1),  463--471 (2008).

\bibitem{Martinache2009}
Martinache, F., Guyon, O., Lozi, J., Garrel, V., Blain, C., and Sivo, G.,
  ``{The Subaru coronagraphic extreme AO project},'' {\em Exoplanets and Disks:
  Their Formation and Diversity, 9-12 March 2009}~{\bf 1158},  329--332 (2009).

\bibitem{Cheetham2015}
Cheetham, A.~C., Kraus, A.~L., Ireland, M.~J., Cieza, L., Rizzuto, A.~C., and
  Tuthill, P.~G., ``{MAPPING the SHORES of the BROWN DWARF DESERT. IV.
  OPHIUCHUS},'' {\em Astrophysical Journal}~{\bf 813}(2),  1--23 (2015).

\bibitem{Dong2012}
Dong, S., Haist, T., Osten, W., Ruppel, T., and Sawodny, O., ``{Response
  analysis of holography-based modal wavefront sensor},'' {\em Applied
  Optics}~{\bf 51}(9),  1318 (2012).

\bibitem{Wilby2016}
Wilby, M.~J., Keller, C.~U., Snik, F., Korkiakoski, V., and Pietrow, A. G.~M.,
  ``{The coronagraphic Modal Wavefront Sensor: a hybrid focal-plane sensor for
  the high-contrast imaging of circumstellar environments},'' ~{\bf 112},
  1--14 (2016).

\bibitem{tuthill2012unlikely}
Tuthill, P.~G., ``{The unlikely rise of masking interferometry: leading the way
  with 19th century technology},'' in [{\em Optical and Infrared Interferometry
  III}{\nolinebreak\hspace{0.1em}]},   {\bf 8445},  844502, International
  Society for Optics and Photonics (2012).

\bibitem{pancharatnam1956generalized}
Pancharatnam, S., ``{Generalized theory of interference, and its
  applications},'' in [{\em Proceedings of the Indian Academy of
  Sciences-Section A}{\nolinebreak\hspace{0.1em}]},   {\bf 44}(5),  247--262,
  Springer (1956).

\bibitem{Berry1987}
Berry, M., ``{The Adiabatic Phase and Pancharatnam's Phase for Polarized
  Light},'' {\em Journal of Modern Optics}~{\bf 34}(11),  1401--1407 (1987).

\bibitem{Escuti:16}
Escuti, M.~J., Kim, J., and Kudenov, M.~W., ``{Controlling Light with
  Geometric-Phase Holograms},'' {\em Opt. Photon. News}~{\bf 27},  22--29 (2
  2016).

\bibitem{Miskiewicz2014}
Miskiewicz, M.~N. and Escuti, M.~J., ``{Direct-writing of complex liquid
  crystal patterns},'' {\em Optics Express}~{\bf 22}(10),  12691 (2014).

\bibitem{Komanduri2013}
Komanduri, R.~K., Lawler, K.~F., and Escuti, M.~J., ``{Multi-twist retarders:
  broadband retardation control using self-aligning reactive liquid crystal
  layers},'' {\em Optics Express}~{\bf 21}(1),  404 (2013).

\bibitem{Kim2015}
Kim, J., Li, Y., Miskiewicz, M.~N., Oh, C., Kudenov, M.~W., and Escuti, M.~J.,
  ``{Fabrication of ideal geometric-phase holograms with arbitrary
  wavefronts},'' {\em Optica}~{\bf 2}(11),  958 (2015).

\bibitem{Doelman2017PatternedSensing}
Doelman, D., Snik, F., Warriner, N., and Escuti, M., ``{Patterned
  liquid-crystal optics for broadband coronagraphy and wavefront sensing},'' in
  [{\em Proceedings of SPIE - The International Society for Optical
  Engineering}{\nolinebreak\hspace{0.1em}]},   {\bf 10400} (2017).

\bibitem{tuthill2010sparse}
Tuthill, P., Lacour, S., Amico, P., Ireland, M., Norris, B., Stewart, P.,
  Evans, T., Kraus, A., Lidman, C., Pompei, E., {others}, and Kornweibel, N.,
  ``{Sparse aperture masking (SAM) at NAOS/CONICA on the VLT},'' in [{\em
  Ground-based and Airborne Instrumentation for Astronomy
  III}{\nolinebreak\hspace{0.1em}]},   {\bf 7735},  77351O, International
  Society for Optics and Photonics (2010).

\bibitem{Norris2015}
Norris, B., Schworer, G., Tuthill, P., Jovanovic, N., Guyon, O., Stewart, P.,
  and Martinache, F., ``{The VAMPIRES instrument: Imaging the innermost regions
  of protoplanetary discs with polarimetric interferometry},'' {\em Monthly
  Notices of the Royal Astronomical Society}~{\bf 447}(3),  2894--2906 (2015).

\bibitem{snik2012}
Snik, F., Otten, G., Kenworthy, M., Miskiewicz, M., Escuti, M., Packham, C.,
  and Codona, J., ``{The vector-APP: a broadband apodizing phase plate that
  yields complementary PSFs},'' {\em Proc. SPIE}~{\bf 8450},
  84500M--84500M--11 (2012).

\bibitem{carlotti2010new}
Carlotti, A. and Groff, T., ``{New approaches to the design of non-redundant
  aperture masks},'' in [{\em Ground-based and Airborne Telescopes
  III}{\nolinebreak\hspace{0.1em}]},   {\bf 7733},  77333D, International
  Society for Optics and Photonics (2010).

\end{thebibliography}
\bibliographystyle{spiebib} 

\end{document}